\documentstyle[12pt]{article}
\def\ben{\begin{enumerate}}  \def\een{\end{enumerate}}
\def\beq{\begin{equation}}   \def\eeq{\end{equation}}
\def\bea{\begin{eqnarray}}  \def\eea{\end{eqnarray}}

\def\lsim{\raise0.3ex\hbox{$<$\kern-0.75em\raise-1.1ex\hbox{$\sim$}}}
\def\gsim{\raise0.3ex\hbox{$>$\kern-0.75em\raise-1.1ex\hbox{$\sim$}}}

\begin{document}

\vbox to 1 truecm {}
\begin{center}
{\large \bf A tentative new approach to the 
Schr\"odinger cat problem
}
\par

\vspace{1 truecm}

{\bf Bernard d'Espagnat}\\
{\it Laboratoire de Physique Th\'eorique}\\
{\it Unit\'e Mixte de Recherche
UMR 8627 - CNRS}\\    {\it Universit\'e de Paris XI,
B\^atiment 210, 91405 Orsay Cedex, France}  
\end{center} 

\vspace{2 truecm} 
\begin{abstract}
Though scientifically unconvincing, the Broglie-Bohm model has the virtue of reproducing the observational predictions of quantum
mechanics while being conceptually crystal-clear. Hence, even if we do not believe in it, we may find it useful in suggesting ways of
removing conceptual difficulties. This procedure is here applied to the Schr\"odinger cat riddle (in its Wigner version). The outcome
yields tentative views on the relationship between mentality and physical reality. 
\end{abstract}

\vskip 2 truecm

\noindent {\it Submitted to Physics Letters A}

\newpage
\pagestyle{plain}

What is it that quantum mechanics describes? Is it ``man-independent reality'' or ``communicable human experience''? Philosophers have
ready-made, but conflicting, answers to this. Physicists have access to more factual elements of appreciation. But even  with their
help, the answer is not very clear. Many of us still consider that the ultimate purpose of physics can only be to ``lift the veil of the
appearances'' and describe reality as it really is. However it is well known that to render the quantum formalism ontologically
interpretable requires either altering it in more or less ad hoc ways (plugging in nonlinear terms in the Schr\"odinger equation) or
assuming explicitly that hidden variables exist. And, for well-known reasons, both procedures are somewhat unpalatable. In view of this,
a number of physicists tend to implicitly endorse a purely instrumentalist conception of physics. This approach in no way amounts to
rejecting as meaningless the notion of a basic man-independent reality, as pure idealist thinkers do. It consists in observing that,
whatever this entity ``really is'', presumably it differs even more than previously thought from what it looks like. And that,
consequently, the more secure standpoint is not to take sides, in an a priori manner, on the question whether or not it should be
pictured and, if yes, how it should be. Incidentally, note in this respect that, within this standpoint, the completeness assumption
should not be bluntly stated as ``hidden variables do not exist'', since this is a metaphysical hypothesis, even though a negative one.
Following Stapp [1], it should be expressed as the assumption that ``no theoretical construction can yield experimentally verifiable
predictions about atomic phenomena that cannot be extracted from a quantum theoretical description''. This leaves open the possibility
that hidden variables exist, provided that they should be ``really hidden''. Otherwise said, as here understood the instrumentalist
conception is still quite general. For example, it is compatible with both pure antirealism and Bohm's ideas, especially when the latter
are expressed via the conception of a real {\it implicit} order, differing very much from the {\it explicit} one, that reflects but the
appearance of reality.\par

	As a consequence of the fact that it postulates so little, the instrumentalist conception has considerable advantages. In particular,
it was recently stressed in these columns [2] that it is essentially within its framework that decoherence can truly be said to solve
the quantum measurement problem. This, of course, is a considerable advantage. However, it remains true that the instrumentalist
conception also raises questions. One of them concerns our final perception of objects (see e.g. [2], Appendix and the there given
references) and is of a very much philosophical nature\footnote{It is well known that, by itself, decoherence does not unambiguously
produce localization. In the instrumentalist conception, to account for what is observed it is necessary to assume, in addition, that
locality is an a priori mode of human sensibility, much as Kant assumed {\it space} to be one.}. It is not to be debated here. But it is
not  the only one. Actually, the fact of considering the quantum formalism as merely being a set of observational predictive rules gives
rise to another conceptual riddle. And indeed what is claimed in the present article is that an analysis of the latter sets us on the
path of a conjecture that seems worthy of attention.\par

	The riddle in question is one of self-consistency and has to do with ``consciousness'' or ``mentality''. Schr\"odinger approached it
when he proposed replacing the measuring apparatus by a cat, for a cat, presumably, has a consciousness of some sort. Wigner [3]
approached it even more when he substituted a friend of him for the cat, for a friend surely is conscious. To explain what it consists
of, let us consider a measurement performed on a system $Q$ not lying in an eigenstate of the measured observable, $A$. Let then $S$ be
the system composed of $Q$, the pointer and the environment ($S$ being considered after the measurement has taken place), and let us
explicitly include Wigner's friend -- together with his mind of course -- within $S$. For simplicity sake we may assume that observable
$A$ has a discrete spectrum. Within a statistical ensemble $E$ of such $S$'s, each one of the friends then sees ``his'' pointer lying in
a graduation interval corresponding to one definite eigenvalue of $A$. In this respect, his knowledge is not -- so it seems --
qualitatively different from the one that an external observer having observed the pointer would have. Apparently, we are therefore
forced to consider that $E$ is a (proper) mixture of $S$ systems on each of which a measurement of $A$ has been performed. However, it is
well known that among the predictions yielded by the density operator attached to such a mixture and that are verifiable {\it in
principle} (assuming the availability of arbitrary large and complex instruments, arbitrary long measurement times etc) there are some
that are incompatible with those directly following from our starting assumption, according to which, before the measurement, $Q$ was not
in an eigenstate of $A$ (see [4] or [5]). In other words, we fall back here on the well-known difficulty of quantum measurement theory.
As explained in [2], decoherence theory removes this difficulty as long as the measuring instruments are considered as inanimate and the
ultimate act of observing is attributed to an impersonal and collective WE, thought of as looking down, so to speak, upon the physical
world. But, as we see, the riddle reappears as soon as we take up a more down-to-earth view on the problem, as Wigner invites us to
do.\par

Now, when faced with problems such as this one -- that are neither mathematical nor purely physical but conceptual -- we, physicists,
feel somewhat at a loss until we find some similarity between them and problems that are, to us, more familiar. In this respect, the old
Louis de Broglie-Bohm hidden variable model (with pilot wave or quantum potential) may be of help. Not that we should necessarily
believe it is true. Many arguments (e.g. those described in detail in [6] and [7]) speak against it. But it does reproduce all the
observational quantum mechanical predictions; it yields, to the quantum measurement problem, a solution differing from the above one but
fully consistent as well\footnote{In it, the representative point is, right from the start, determined to proceed  into one or the
other of the sectors of configuration space corresponding to the possible pointer positions.}; and it has the great advantage of being
conceptually crystal-clear. It can therefore be used as a theoretical laboratory. If, within it, it proves possible to take explicitly
into account the (undeniable!) fact that WignerÕs friend is conscious, it is conceivable that the basic idea underlying this solution
can be extended, outside the model, to the general theory. Incidentally, note that in the model the above-mentioned difference between
implicit and explicit order of course holds. The implicit order concerns the hidden variables that, together with the nonlocal
pilot-wave, compose {\it man-independent reality}. The explicit order is the order that is manifest in the appearances that compose the
set of the observed phenomena, alias {\it empirical reality}.\par

	According to the model, within a Young-type thought-experiment with two slits the particle is driven by a pilot-wave that passes
through both slits at once; and this has the consequence that, in the model, while each particle passes through but one slit, fringes
nevertheless appear. We can therefore say that, in the model, the particle is at any time at some well-defined place even though it
takes part in a typically quantum phenomenon. In this respect it resembles the friend in WignerÕs parable, who is at any time in some
well-defined state of consciousness while he also is taking part in a quantum phenomenon. To strengthen the analogy it is then
appropriate that, in the model, we should attribute to the particle some kind of a mentality (or, say, protomentality, the physical
nature of which needs not be specified in detail). Within the experiment in question, each one of the involved particles then
``observes'' that, at a certain time, it passes through one, well specified, slit. This is an {\it internal state of consciousness} of
the particle and since, in the model, the particle position is an element of man-independent reality, this internal state of
consciousness has also to be considered as being an element of man-independent reality. For the particle this internal state of
consciousness has no predictive power, since what will happen to the said particle is entirely governed by the pilot wave.\par

	Now, it is often, and quite rightly, said that the very fact of knowing through which slit the particle passes prevents the fringes
from being formed. At first sight this might seem to constitute a valid objection against any idea of attributing a state of
consciousness to the particles. And in fact, so it would be if we assumed that the particle could communicate its knowledge to the world
at large. So let us assume that it can't; that, at least as far as particles and other micro-systems are concerned, ``internal states of
consciousness'' are really private (such an idea is quite in agreement with the fact that, in the model, the ``consciousness state'' in
question is a ``hidden variable'' just as the position itself, since hidden variables do not act on the pilot wave). This being the case,
an external observer $O$, even if we assume he knows of the {\it existence} of such internal states of consciousness, must not take this
existence into account when predicting what will be observed. He therefore predicts the fringes will appear, in conformity with
experiment.\par

	On the other hand, imagining a category of states of consciousness that would always remain totally hidden would obviously be quite
pointless. Hence, we should ask whether circumstances exist in which statements bearing on such ``interior'' states may, after all, have
some relationship with the public domain of shared experience (while remembering of course that this domain is the one of the
{\it explicit} order, that is, in the model, the one of ``appearances that are the same for everybody''). Now, decoherence helps us here.
For, in the Young-type experiment, call $\psi_1$ and $\psi_2$ the partial wave functions issuing from slits 1 and 2 respectively, and
suppose we replace the micro particles by corpuscles that are appreciably larger and whose interaction with the environment is,
consequently, not negligible. The fringes then fade and, when the corpuscles are macroscopic enough, they practically disappear. For the
purpose of predicting outcomes of future observations, the ensemble of the involved corpuscles can then be treated as a mixture of two
``pure cases'' described by the wave functions $\psi_1$ and $\psi_2$. Now $\psi_1$ ($\psi_2$) is just the wave function that an observer
would attribute to a set of corpuscles known to have passed through slit 1 (2). This shows that, in such circumstances, the internal
state of consciousness of the corpuscles passing through one particular slit may indeed be considered without harm as having the
intersubjective predictive role normally attributed, in quantum mechanics, to pieces of knowledge obtained from measurements.
Incidentally, note that this reasoning is fully consistent with the Broglie-Bohm model since, as far as mere predictions are concerned,
this model yields the same ones as non-relativistic quantum mechanics. \par

	If we now turn back to the Wigner friend problem and still consider it within the Broglie-Bohm model, we find that the foregoing views
can quite naturally be fitted to it. Indeed, for the same reasons as above, we can assume without inconsistency that $S$ has an internal
state of consciousness as long as we do not attribute predictive power to it, and we can also relax somewhat this condition when we take
into account the fact that $S$ is macroscopic and interacts therefore with its environment. In fact, for the purpose of predicting what
will {\it practically} be observed, (forgetting about measurements that are conceivable only in principle) an ensemble $E$ of thus
prepared $S$'s can be treated as a mixture, for decoherence is at work. And -- just as above -- the quantum states composing this mixture
are the ones that the various possible internal states of consciousness of the friends would generate if these consciousness states were
viewed as predictive. Hence, within the ``set of observational predictive rules'' that quantum mechanics merely is (in our
instrumentalist approach), these internal states of consciousness can be considered as representing elements of {\it empirical} reality.
More precisely they can be viewed as coinciding in nature with the predictive states of consciousness that we refer to when we state that
such and such a measurement outcome has been observed.\par

	This being the case, we can now revert to the general theory. We can consider the Broglie-Bohm model as just being one conceivable
description of man-independent reality, drop the assumption that it does picture it as it really is, but, still, try to make good use of
the suggestion this model provided us with. Clearly, the essential element in the latter is the notion of {\it internal states of
consciousness}. In the model such ``states'', existing at the elementary level, are, as we noted, elements of man-independent reality,
and this is a feature that definitely should be kept. In the general case they, correspondingly, are devoid of any predictive power,
either internal (for the systems themselves) or external (to the public at large, knowing they exist makes no difference).  Hence, in a
Young-type experiment, their existence is fully compatible with the building up of the fringes.\par

	On the other hand, it is also true that, in a Young-type experiment supposedly made with macroscopic ``thinking corpuscles'', the latter
could, because of decoherence, make use of a wave function matched to their internal states of consciousness in order to predict their
own future behavior. And so can, of course the friends in a Wigner-friends type of setting. Moreover, still because of decoherence, in
both cases the external observer $O$ makes no false prediction if, for the purpose of calculating what he will observe when making
{\it practically} feasible measurements, he chooses to take the existence of the said states of consciousness seriously into
account.\par

	To sum up, within this conception it is considered that even microsystems can be endowed with ``internal states of consciousness'' (or
``protoconsciousness'', whatever this may be) that are elements of a basic, not publicly accessible, reality, rather than of empirical
reality. In other words, they are hidden (remember we left open the possibility that hidden variables should exist, provided that they
should be ``really hidden''). It is only when the involved systems become macroscopic enough for their interaction with the environment
to be appreciable that these internal states of consciousness obtain some degree of public significance. This means that they gain
predictive power. It then turns out that the impressions they predict can usually be described in a realist {\it language}, that is,
{\it as if} they referred to objects existing per se. The set of such intersubjective appearances is what is called here ``empirical
reality''. It is thus meaningful to speak of a kind of ``co-emergence'' of ``public'' states of consciousness on the one side and
empirical reality on the other side, out of a  ``man-independent reality'' that, itself, lies beyond our intersubjective describing
abilities. \\

\noindent {\it Discussion} \par

	It goes without saying that the conception of the relationships between mind and reality we arrived at is extremely sketchy and
speculative. In spite of this, it is interesting to note that, surprisingly enough, it shows quite a definite similarity with
Whitehead's views: For indeed the notion of internal states of consciousness is quite akin to the ones of ``prehensions'' and
``occasions of experience'' that, in Whitehead's philosophy, play a basic role even at the elementary particle level. \par

In this, it is quite at variance with the ``received view'', according to which only things with highly complex structures have
consciousness. This received view considers consciousness as being, somehow, an attribute of such things. This means, it implicitly
postulates that our ``theory of things in general'' (physics) should be ontologically interpretable, for if things are mere
``phenomena'', that is, ``appearances to consciousness'', how could consciousness be an attribute of things? Now, there does exist
ontologically interpretable models and it may well be, after all, that one or another of them is right. And since these models yield an
elementary solution to the Wigner friend riddle\footnote{In the spontaneous collapse models, for instance, the collapse spontaneously
taking place in $S$ has the effect that even measurements made on systems involving the environment could not -- not even in principle!
-- be inconsistent with the predictions yielded by the view that the $S$'s are in eigenstates of $A$, as WignerÕs friend sees them to
be.}  a supporter of any one of them is fully entitled to claim that this ``received view'' constitutes the most reasonable way of
introducing consciousness into our picture of the World. The approach described in this article can thus be of interest only to the
physicists who, as explained here in the beginning, are reluctant to adhere to any one of these models and therefore explicitly or
implicitly take up an instrumentalist standpoint.\par
  
	It is suggested here that, for them, this interest is twofold. The first interest may be labeled ``technical''. It consists in that the
instrumentalist conception was shown above to be free from the ``remaining''\footnote{``Remaining'', even after it has been checked that,
due to decoherence, the quantum measurement theory is free from contradiction as long as the various parts of the measuring instruments
are considered inanimate.} lack of self-consistency of which it seemed to suffer in connection with the Wigner friend riddle and the
consciousness of the said friend\footnote{Admittedly, we knew a priori that this difficulty is only an apparent one. The reason is that
the Broglie-Bohm model exactly reproduces all the experimentally verifiable predictions of non relativistic quantum mechanics and is
thus a concrete example of how we could consistently imagine that the predictions constituting the instrumentalist conception get
materialized. The fact that this model does remove the difficulty (see note 2) then shows that, within the instrumentalist conception,
the difficulty in question can somehow be overcome. But of course it is still nicer to know {\it how} it can be.}. The second one is of
a more general nature. It is related to the importance of the role that the notion of consciousness -- more precisely, that of ``getting
aware'' -- has taken up both in quantum physics (due to the intersubjective character of the Born rule) and in the XXth century
epistemology (to the extent that the latter adopted Schlick's axiom that the meaning of a statement {\it boils down} to its method of
verification: for any verification procedure ultimately supposes the existence of some conscious verifier).  When a theory thus imparts
a key role to a given concept and, at the same time, leave this concept unanalyzed and hazy, this is certainly a weakness it has. The
content of the present article may be viewed as a step in the direction of curing this weakness. Up to now, neurobiological and
philosophical investigations with the same end were essentially carried out within a purely classical representation of the World. While
such a procedure can be justified on the basis of significant arguments, it remains that within an (explicit or implicit) purely
instrumentalist approach it is open to criticism since (to repeat) it is objectionable to claim, as pure instrumentalist scientists
sometimes tend to do, both that consciousness is a property of things and that, ultimately, things are mere appearances... to
consciousness. By making internal states of consciousness elements of basic reality, by distinguishing them from the communicable states
of consciousness (which belong to empirical reality) and by linking these two notions together (in the case of macro-systems) with the
help of decoherence, the conception here arrived at removes the very basis of this criticism and, at the same time, provides us with a
not totally hazy picture of consciousness. It may therefore be worth considering.

\end{document}